\documentclass[preprint2]{aastex}

\shorttitle{Invariant plane and planet 9}
\shortauthors{Gomes et al.}

\begin{document}


\title{The inclination of the planetary system relative to the solar equator may be explained by the presence of Planet 9}

\author{Rodney Gomes}
\affil{Observat\'orio Nacional \\
Rua General Jos\'e Cristino 77, CEP 20921-400, Rio de Janeiro,  Brazil}
\email{rodney@on.br}

\author{Rogerio Deienno}
\affil{Laboratoire Lagrange, UCA, OCA, CNRS, Nice, France}
\affil{Instituto Nacional de Pesquisas Espaciais, S\~ao Jos\'e dos Campos, SP, Brazil}

\author{Alessandro Morbidelli}
\affil{Laboratoire Lagrange, UMR7293, Universit\'e C\^ote d’Azur, CNRS, Observatoire de la C\^ote d’Azur}

\begin{abstract}
We evaluate the effects of a distant planet, commonly known as planet 9, on the dynamics of the giant planets of the Solar System. We find that, given the large distance of planet 9, the dynamics of the inner giant planets can be decomposed into a classic Lagrange-Laplace dynamics relative to their own invariant plane (the plane orthogonal to their total angular momentum vector) and a slow precession of said plane relative to the total angular momentum vector of the Solar System, including planet 9. Under some specific configurations for planet 9, this precession can explain the current tilt of $\sim 6^{\circ}$ between the invariant plane of the giant planets and the solar equator. An analytical model is developed to map the evolution of the inclination of the inner giant planets' invariant plane as a function of the planet 9's mass, inclination, eccentricity and semimajor axis, and some numerical simulations of the equations of motion of the giant planets and planet 9 are performed to validate our
analytical approach.  The longitude of the ascending node of planet 9 is found to be linked to the longitude of the ascending node of the giant planets' invariant plane, which also constrain the longitude of the node of planet 9 on the ecliptic. Some of the planet 9 configurations that allow explaining the current solar tilt are compatible with those proposed to explain the orbital confinement of the most distant Kuiper belt objects. Thus, this work on the one hand gives an elegant explanation for the current tilt between the invariant plane of the inner giant planets and the solar equator and, on the other hand, adds new constraints to the orbital elements of planet 9.

\end{abstract}


\section{Introduction}

The gradual discovery of increasingly distant trans-Neptunian objects (TNOs) has allowed new tests for the existence of a yet undiscovered distant planet in the solar system. \cite{gomesetal2015} analyzed the large semimajor axis centaurs and concluded that they are produced continually by the decrease of perihelia of scattered disk objects, induced by the perturbation of a distant planet. \cite{trujillo2014}, as they announced the discovery of the distant TNO 2012 VP113, also noted that distant TNOs not perturbed by close encounters with Neptune show a remarkable alignment of their arguments of perihelia and proposed that a distant planet is responsible for this alignment. More recently, \cite{BaBr2016} studied more deeply the orbital alignment of those distant TNOs, showing that the six most distant objects exhibit also a clustering in their longitudes of node; they estimated that the probability that this double alignment in argument of perihelion and longitude of the node is just fortuitous is $0.007$\%. Moreover they showed that a planet 9 (hereafter named just pl9) could account for said alignment if it had a mass of about $10 M_{\oplus}$ and an orbit with semimajor axis between $300$ and $900$ au, perihelion distance between $200$ and $350$ au, and orbital inclination of about $30^{\circ}$ to the ecliptic plane. The Batygin-Brown approach based on secular dynamics is able to determine an approximate orbit for the distant planet that could explain the said alignment, but not the planet's position on that orbit. \cite{fienga2016} use a typical orbit among those proposed by \cite{BaBr2016} and  determined the range in true longitude of pl9 on that orbit that decreases the residuals in INPOP ephemerids of Saturn, relative to the Cassini data. \cite{hol-payne2016} obtained a similar result using JPL ephemerids. \cite{BrBa2016} refined their previous results by further constraining the mass and orbital elements of pl9 that are compatible with the observed TNOs orbital alignment. They now argue for pl9's semimajor axis in the range $380–980$au, perihelion distance in the range $150–350$ au and a mass between $5$ and $20 M_{\oplus}$, for an orbital inclination of $30^{\circ}$. \cite{malhotra2016} looked for extra constraints on pl9 orbit by analyzing the orbital periods of the four longest period TNOs. Their approach is based on the supposition that pl9 is in mean motion resonances with those TNOs. \cite{beust2016}, however, showed that a mean motion resonant configuration is not necessary to explain the orbital confinement. 

Here we study the precession of the plane orthogonal to the total angular momentum of the four giant planets due to the perturbation of pl9.  We find that, given the large distance of pl9, the dynamics of the giant planets can be decomposed into a classic Lagrange-Laplace dynamics relative to their own
invariant plane (the plane orthogonal to their total angular momentum vector, hereafter named iv4) and a slow precession of said plane relative to the total angular momentum vector of the Solar System, including pl9. Planetary system formation predicts that planets are formed from a disk of gas and dust and this disk rotates on the same plane of the star's equator. The final planetary orbits, if no mutual close encounters take place, must be approximately coplanar and coincident with the star's equator. We thus suppose that the giant planets and the solar equator were initially on the same plane. We assume that pl9 was scattered away from the region of the other giant planets when the disk was still present and the solar system was still embedded in a stellar cluster \citep{izidoroetal2015}. The stellar cluster is needed, so that the perihelion of the orbit of pl9 can be lifted and pl9 can decouple from the other planets \citep{brasseretal2008}. Because most of the angular momentum is in the protoplanetary disk, it is likely that the ejection of pl9 onto an inclined orbit did not significantly change the inclination of the disk and of the other giant planets. Notice also that the inclination of pl9 might have been increased by the action of the cluster, while lifting the perihelion in a Lidov-Kozai like dynamics \citep{brasseretal2008}. Thus, we assume that, at the removal of the protoplanetary disk and of the birth cluster of the Sun, the 4 major giant planets were on orbits near the solar equator, while pl9's orbit was off-plane. At this point, a slow precession of iv4 started to take place relative to the total angular momentum vector of the Solar System, including pl9, keeping however the orientation of the solar equator plane unchanged. Thus the current angle between the solar equator and iv4 (about $6^{\circ}$ - see below) must be a signature of pl9 perturbation and we aim at finding ranges of orbital elements and mass for pl9 that can explain quantitatively the present tilt of iv4 relative to the solar equator.

The solar equator with respect to the ecliptic is identified by an inclination $I_S=7.2^{\circ}$ and a longitude of the ascending node $\Omega_S=75.8^{\circ}$ \citep {beck&giles}. The invariant plane with respect to the ecliptic is defined by an inclination $I_i=1.58^{\circ}$ and a longitude of the ascending node $\Omega_i=107.58^{\circ}$ \citep {souami2012}. Employing two rotations we can find the invariant plane  angles with respect to the solar equator to be $I_v=5.9^{\circ}$ and $\Omega_v=171.9^{\circ}$. We will use these parameters throughout the rest of the paper. We also consider that iv4, as above defined, is equivalent to the invariant plane of the solar system mentioned in the works above, which take into account also the inner planets.

In Section 2 we develop two analytical approaches aimed at determining the tilt experienced by iv4 due to the perturbation of pl9. We also perform some numerical integrations of the full equations of motion to validate our analytical approaches. In Section 3, we apply our analytical method to determine the range of masses and orbital elements of pl9 that can account for the observed tilt of iv4 to the solar equator plane. In Section 4, we draw our conclusions.

\section{  Methods }

We first apply the classical Laplace-Lagrange formalism up to second order in the inclination to evaluate the variation of the inclination experienced by iv4 to due pl9. Since a second order approach may not be sufficient for large inclinations of pl9, we develop another approach based on the angular momenta of pl9 and the four giant planets. In this case, we make no approximation on the inclinations but just a first order approximation in the ratio of the known planets' semimajor axes to that of pl9. 

\subsection{First approach: secular perturbations to second order}

Following \citet{batyginetal2011}, we derive a secular theory of the evolution of the inclination of iv4 based on the classical Laplace-Lagrange theory up to the second order in the inclinations. From \cite{murray-dermott1999}, we have the following form for the classical Hamiltonian: 

\begin{equation}
H = \frac{1}{2}\sum_{j=1}^N\sum_{k=1}^N B_{jk}I_jI_k\cos(\Omega_j - \Omega_k)
\label{murray-hamiltonian}
\end{equation}  

\noindent where $j$ and $k$ indicate the perturbed and the perturbing bodies respectively. The $Nth$ index refers to pl9. All inclinations are expressed with respect to the solar equator, supposed as the initial fixed reference frame. The coefficients $B_{jj}$ and $B_{jk}$ assume the form:

\begin{eqnarray}
B_{jj} &=&  -\frac{n_j}{4} \sum_{k=1,k\ne j}^N \frac{m_k}{M_\odot+m_j}\alpha_{jk}\bar{\alpha}_{jk}b_{3/2}^{(1)}(\alpha_{jk}) \nonumber \\
B_{jk} &=& \frac{n_j}{4}\frac{m_k}{M_\odot+m_j}\alpha_{jk}\bar{\alpha}_{jk}b_{3/2}^{(1)}(\alpha_{jk})
\label{murray-bjk}
\end{eqnarray}

\noindent where $M_\odot$ is the mass of the Sun, $m_j$ and $m_k$ are the masses of the  interacting bodies, and $n_j$ is the mean motion of the planet $j$. $\alpha_{jk} = a_j/a_k$ and $\bar{\alpha}_{jk} = \alpha_{jk}$ for $a_j<a_k$. For $a_j>a_k$ we have $\alpha_{jk} = a_k/a_j$ and $\bar{\alpha}_{jk} = 1$. $b_{3/2}^{(1)}(\alpha_{jk})$ is the Laplace coefficient of the first kind \citep[Ch. 7]{murray-dermott1999}.

In the context of the Laplace-Lagrange secular theory, valid for small values of the inclination and eccentricity, if we want to account for large values of the inclination and eccentricity of pl9 we need to add some new ingredients to the classical theory. In this manner, the inclination of pl9 was accounted for by reducing pl9's mass by a factor of $\sin I$, i.e. ${m_9}_{(new)}={m_9}_{(real)}\cos I_9$. By doing this, we consider only the projection of the mass of pl9 onto the planet's reference frame (iv4) \citep{batyginetal2011}.
As for the eccentricity of pl9, assuming that one cannot derive a simple secular approach \citep{murray-dermott1999}, it turns necessary to somehow incorporate the averaged effect of an eccentric orbit upon the motion of the perturbed planet. According to \citet{gomesatal2006}, the averaged effect can be computed assuming that the perturber is on a circular orbit of radius $b$, where $b_9=a_9\sqrt{1-e_9^2}$ is the semi minor axis of the real perturber's orbit. Thus, in order to compute for the possible large eccentricity of pl9 we will assume $b_9=a_9\sqrt{1-e_9^2}$ as its circular semimajor axis analog.

Therefore, with the implementations of ${m_9}_{(new)}={m_9}_{(real)}\cos I_9$ and $b_9=a_9\sqrt{1-e_9^2}$, it is possible to rewrite the Hamiltonian (\ref{murray-hamiltonian}) in terms of the vertical and horizontal components of the inclination ($p_j=I_j\sin \Omega_j$ and $q_j=I_j\cos \Omega_j$), where the first-order perturbation equations ($\dot{p}_j=\partial H_j/\partial q_j$ and $\dot{q}_j=-\partial H_j/\partial p_j$) lead to an eigensystem that can be solved analytically \citep[Ch. 7]{murray-dermott1999}

\begin{eqnarray}
p_{j} &=& \sum_{k=1}^N I_{jk}\sin(f_kt+\gamma_k) \nonumber \\
q_{j} &=&  \sum_{k=1}^N I_{jk}\cos(f_kt+\gamma_k)
\label{murray-pq}
\end{eqnarray}
\noindent with $f_{k}$ being the set of $N$ eigenvalues of matrix {\bf B} (Eq. \ref{murray-bjk}), $I_{jk}$ the associated eigenvector, and $\gamma_k$ a phase angle determined by the initial conditions. This leads to the final solution

\begin{eqnarray}
I_{j} &=& \sqrt{p_j^2 + q_j^2} \nonumber \\
\Omega_{j} &=&\arctan(\frac{p_j}{q_j}).
\label{murray-solve}
\end{eqnarray}

Finally, starting with Jupiter, Saturn, Uranus, and Neptune in the equatorial plane of the Sun, given the orbital parameters of pl9, one can verify that the eigenvectors ($I_{j5}, j=1...4$) have the same magnitude. In this way, the equations in (\ref{murray-solve}) also represent the evolution of the pair (I,$\Omega$) of iv4.
Despite the modifications we introduced to account for the large inclination and eccentricity of pl9, our method has limitations, being less accurate for large values of $I_9$ and $e_9$.

\subsection{Second approach: angular momentum}

The equation of motion of a planet around a star perturbed by a second planet in a reference frame centered in the star is: 

\begin{equation} \label{newton}
\ddot {\vec {r}} = - G (m_9 + M) \frac{{\vec r}} {{r^3}} + G m_9 \left\{ \frac{{\vec r_{19}}} {{r_{19}^3}} - \frac {{\vec r_9}} {{r_9^3}} \right\}
\end{equation}

\noindent where the subscript $9$ refers to the perturbing planet and the perturbed planet has no subscript. In this equation, $\vec r$ is the radius vector and $r$ its absolute value, $m_9$ is the perturbing planet mass, $M$ the star's mass, $G$ the gravitational constant and $r_{19}$ the distance between both planets. Let us define the angular momentum per unit mass by $\vec{h} = \vec{r} \times \dot{\vec{r}}$. Using Eq. \ref{newton}, the time derivative of $\vec h$ can be found as:

\begin{equation} \label{ang-mom}
\dot {\vec {h}} = \left\{ G m_9 \frac { {\vec r \times \vec r_{19}}} {{r_{19}^3}} - G m_9 \frac {{\vec r \times \vec r_9}} {{r_9^3}}\right\}
\end{equation}

Since $\vec r_{19} = \vec r_9 -\vec r$ and $\vec r \times \vec r = 0$, we have to deal just with the vectorial product $\vec r \times \vec r_9$ in Eq. \ref{ang-mom}. We now want to average the right hand of Eq. \ref{ang-mom} in the fast variables for both planets. For that we suppose two reference frames defined on each of the planets' orbits making an angle $I$ between them. The frames are defined by $(\vec i, \vec j, \vec k)$ and $(\vec i_9, \vec j, \vec k_9)$, unitary vectors where the component $\vec j$ is common to both frames. $\vec j$ is in the intersection of the orbital planes and lies on the invariant plane defined by both planets; $\vec i$ and $\vec i_9$ are orthogonal to $\vec j$ on each of the orbital planes and $\vec k$ and $\vec k_9$ completes the reference frames through the right hand rule. It must be noted that these frames are defined just to compute the derivative of $\vec h$ on those components instantaneously. We now assume that the perturbed planet has a small enough eccentricity so as to consider its orbit as circular. On the other hand, the perturbing planet will be considered eccentric. In this manner we can represent the radius vector of each of the planets as:

\begin{equation} \label{ijk}
\vec r = (a \cos{l})\; \vec i + (a \sin{l})\; \vec j \\
\end{equation}

\begin{equation} \label{ijk9}
\vec r_9 = (r_9 \cos{\theta_9})\; \vec i_9 + (r_9 \sin{\theta_9}) \; \vec j \;\; ,
\end{equation}

\noindent where $a$ is the perturbed planet semimajor axis, $l$ is the perturbed planet mean longitude and $\theta_9$ is the angle from the intersection of the planes to the perturbing planet's position, which is the sum of pl9's true anomaly $f_9$ and the longitude of the ascending node with respect to the invariant plane.

We now put together Eq. \ref{ang-mom}, \ref{ijk} and \ref{ijk9} and develop the vectorial products remembering that $\vec i \times \vec i_9 = - \sin{I} \vec j$, $\vec i \times \vec j = \vec k$, $\vec j \times \vec i_9 = - \vec k_9 = - \sin{I} \vec k + \cos{I} \vec i$ and $\vec j \times \vec j = 0$. Developing the components in $\vec i$ and $\vec k$, we notice that there is always a trigonometric function in at least one of the fast angles to an odd power, which results in a null average for these components.

For the $\vec j$ component, after developing the vectorial product to the first order in $a/r_9$, we obtain:

\begin{equation} \label {firstorder}
(\vec r \times \vec r_9) / r_{19}^3 = - (a \; r_9 \cos{l} \cos{\theta_9} \sin{I})\;  r_9^{-3} \; T
\end{equation}

\noindent where,

\begin{displaymath}
T = 1- \frac{3}{2} \frac{a^2}{r_9^2} + 3 \frac{a}{r_9} \; (\sin{l} \sin{\theta_9} + \cos{l} \cos{\theta_9} \cos{I}) \\
\end{displaymath}

The first order approximation can be quite accurate when $a/r_9$ is small, which is the case of a distant planet perturbing a close in one. Averaging in the fast angles $l$ and $f_9$, which appears in $\theta_9$ and $r_9$, for one orbital period, we arrive at:

\begin{equation} \label{average}
\frac{1}{P P_9} \int_0^P \int_0^{P_9} \frac {\vec r \times \vec r_9} {r_{19}^3} \; dt^2 = \frac {3} {8} a^2 b_9^{-3} \sin{2 I}
\end{equation}

\noindent where $P$ and $P_9$ are the orbital periods of the perturbed and perturbing planets, respectively, and $b_9= a_9 \sqrt{(1-e_9^2)}$ is the semiminor axis of the perturbing planet. The term $(\vec r \times \vec r_9) / r_9^3$ averages to zero in all components. The variation of $\vec h$ becomes:

\begin{equation} \label{freq}
\dot {\vec h} =  \frac {3} {8} G \,m_9\, a^2 b_9^{-3} \sin{2 I} \vec j
\end{equation}

Through the way the reference frames were constructed, we have shown that $\vec h$ has only and always a non-zero time derivative in the direction of the intersections of the orbital planes. Since the choice of the reference frames could be for any time, we conclude that at any time the non-zero component of the time derivative of $\vec h$ is orthogonal to $\vec h$ on the intersection of the orbital planes. This is satisfied only if the projection of $\vec h$ on the invariant plane is a circle around the origin. The radius of the circle $\alpha$ (where $\sin{\alpha} = H_9/H_t \sin{I}$) is the angle between $\vec h$ and $\vec H_t$ where $\vec H_t = m \vec h + \vec H_9$ is the total angular momentum and $\vec H_9 = m_9  \vec r_9 \times \dot{\vec r_9}$ is pl9 angular momentum. The precession frequency is the coefficient multiplying $\vec j$ in Eq. \ref{freq} divided by $2 \pi \alpha h$, where $h$ is the absolute value of $\vec h = \sqrt{G M_{\odot} a}$.

This approach to compute the variation of the inclination and node of a planet perturbed by another one can be extrapolated to the case of a distant planet perturbing several close in planets. We noticed by numerical integrations that this approach is accurate enough for the Solar System giant planets perturbed by a distant planet, when we replace the four planets by only one with semimajor axis at $10.227$ au and the same angular momentum as the resultant angular momentum of the giant planets.

\begin{figure}
\includegraphics*[clip=true,scale=0.8]{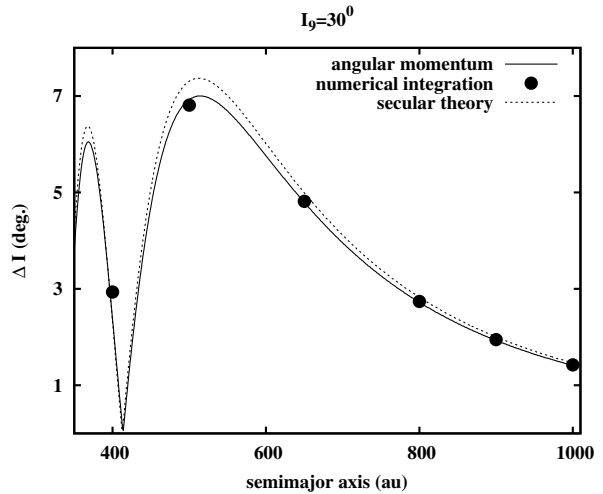}
\caption{The inclination of iv4 relative to the initial plane (the solar equator) after $4.5$ Gy due to the presence of pl9 with $e_9=0.7$, as a function of pl9's semimajor axis. The dashed curve shows the prediction from the secular theory described in sect. 2.1 and the solid curve from the analytic theory described in sect. 2.2. The dots show the results of 6 numerical simulations that contain no approximations.}
\label{compar1}
\end{figure}

\begin{figure}
\includegraphics*[clip=true,scale=0.8]{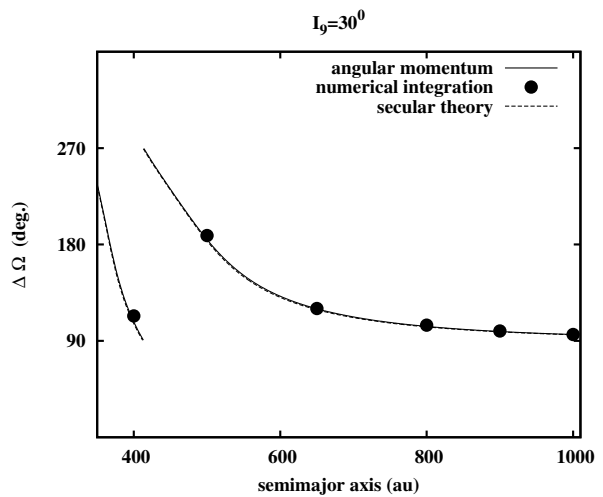}
\caption{the same as Fig. \ref{compar1} but for the longitude of the node of iv4 relative to the solar equator. Here the two analytic approaches look indistinguishable}
\label{compar2}
\end{figure}

\subsection{Comparison of both approaches with numerical integrations}

We considered the current orbital elements of the four giant planets and a pl9 and ran a numerical integration of the full equations of motion for $4.5 Gy$. We noticed that iv4 behaved the same way as if all planets started on the same plane. Thus for simplification we started new integrations with the current giant planets orbital elements except for the inclinations which were started at zero and these integrations were used for the comparisons below. We ran a total of six different numerical integrations with different semimajor axes for pl9 \footnote{We also ran one numerical integration with all eight planets and confirmed that the orbital plane of the inner ones just precessed around a common plane, which in this case would be a iv8 very close to iv4.}. The planet's mass and other orbital elements are $m_9=3 \times 10^{-5} M_{\odot}$, $e_9=0.7$, $I_9=30^{\circ}$, $\Omega_9=113^{\circ}$, $\omega_9=150^{\circ}$ (\cite {BaBr2016}). Figures \ref{compar1} and \ref{compar2} shows the comparison of the two analytic approaches with the numerical integrations. We notice good agreement, thus, from now on we will consider the angular momentum approach to make our analysis.

\begin{figure}
\includegraphics*[clip=true,scale=0.8]{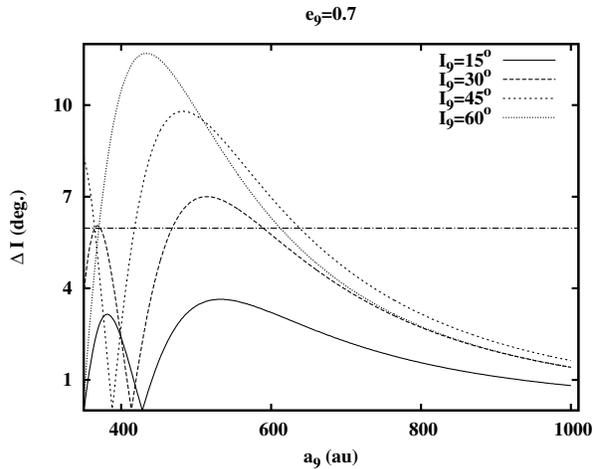}
\caption{Inclination gained by iv4 with respect to the initial reference frame supposed to coincide with the current solar equator for different semimajor axes and initial inclinations of pl9 assuming a mass of $3 \times 10^{-5} M_{\odot}$ and an eccentricity of $0.7$. The horizontal line stand for the current inclination of iv4 with respect to the solar equator.}
\label{varI9}
\end{figure}

\begin{figure}
\includegraphics*[clip=true,scale=0.8]{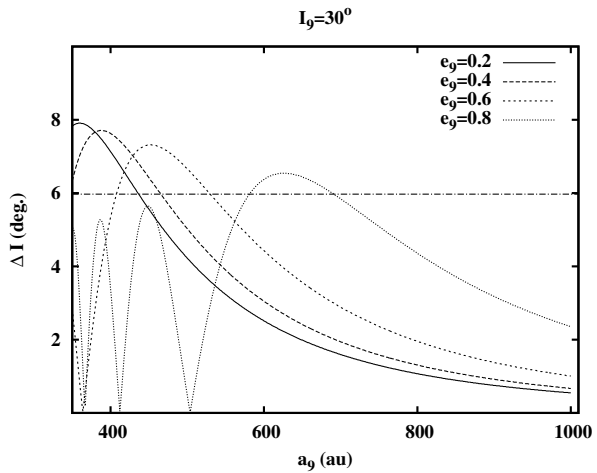}
\caption{The same as Fig. \ref{varI9}, but for different eccentricities and semimajor axes of pl9, assuming an inclination of $30^{\circ}$.}
\label{vare9}
\end{figure}

\section {Constraining a planet 9 that yield $5.9^{\circ}$ tilt}

\begin{figure}
\includegraphics*[clip=true,scale=0.8]{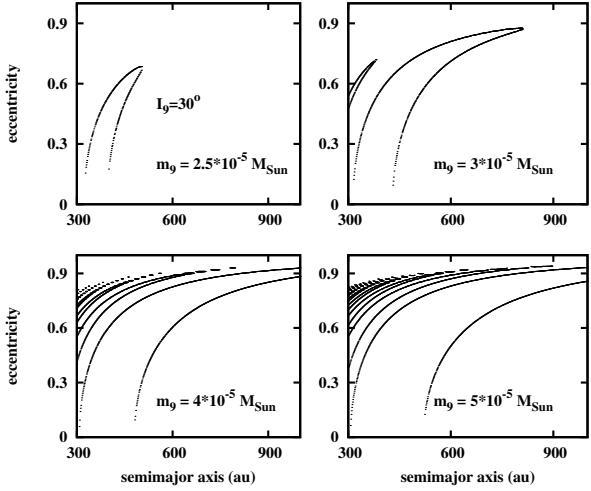}
\caption{Semimajor axis vs. eccentricity of pl9s that yield a tilt of $5.9^{\circ}$ to iv4 with respect to the solar equator, for $I_9=30^{\circ}$}
\label{ae30}
\end{figure}

\begin{figure}
\includegraphics*[clip=true,scale=0.8]{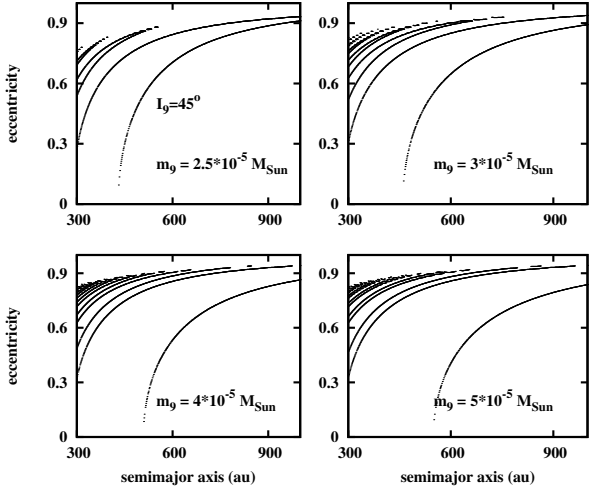}
\caption{The same as Fig. 5 but for $I_9=45^{\circ}$}
\label{ae45}
\end{figure}

Figures \ref{varI9} and \ref{vare9} show how the inclination of iv4 after $4.5$ Gy depends on orbital elements of the perturbing planet. Here we fixed the  mass of pl9 at $3 \times 10^{-5} M_{\odot}$. The largest value of $\Delta I$ in each of these figures stands for the case where the angular momenta of pl9 and iv4 turn $180^{\circ}$ around one another. For smaller semimajor axes of pl9, more than one cycle is accomplished by the pair of angular momentum vectors.  

The plots in Figs. \ref{varI9} and \ref{vare9} allow us to compute pl9 parameters that yield $\sim 5.9^{\circ}$  inclination between the solar equator and iv4 after $4.5$ Gy. for instance, Fig. 3 reveals that a planet of $3 \times 10^{-5} M_{\odot}$, eccentricity of $0.7$ and semimajor axis of $600$ au has to have an initial inclination of $30^{\circ}$ relative to the solar equator to cause the observed tilt.  Figures \ref{ae30} and \ref{ae45} show the loci of pl9 semimajor axis and eccentricity that yield a tilt of $5.9^{\circ}$ of iv4 relative to the solar equator for four possible masses for the distant planet and two initial values of pl9 inclination. We notice that for a mass $2.5 \times 10^{-5} M_{\odot}$ and $I_9=30^{\circ}$, there are just a few choices of planets that can yield a $5.9 ^{\circ}$ tilt. A mass as small as $2 \times 10^{-5} M_{\odot}$ is unable to yield the right tilt for iv4 if $I_9=30^{\circ}$. For higher inclinations of pl9 it is possible for iv4 to achieve a tilt of $5.9^{\circ}$ for somewhat smaller masses of the perturber. The constraints on pl9 orbit that we obtain here have similarities with those obtained by \cite{BaBr2016} using considerations on the orbital alignment of TNOs, although we usually determine a higher eccentricity for a given semimajor axis. For example, the standard pl9 in \cite{BaBr2016} with $m_9=10 M_{\oplus}$ and $I_9=30^{\circ}$ has $a_9=700$ au and $e_9=0.6$. In our case, for the same $m_9$, $I_9$ and $a_9$ the eccentricity must be $0.8$. In \cite{BrBa2016} the best pl9 for $I_9=30^{\circ}$ and $m_9=10 M_{\oplus}$ has $a_9=600$ au and $e_9=0.5$. In our analysis for the same $I_9$, $m_9$ and $a_9$ the eccentricity must be $0.71$. Comparing with \cite{malhotra2016} our eccentricities for pl9 are much higher, since  \cite{malhotra2016} determined eccentricities lower than $0.4$ for $a_9=665$ au.

\begin{figure}
\includegraphics*[clip=true,scale=0.8]{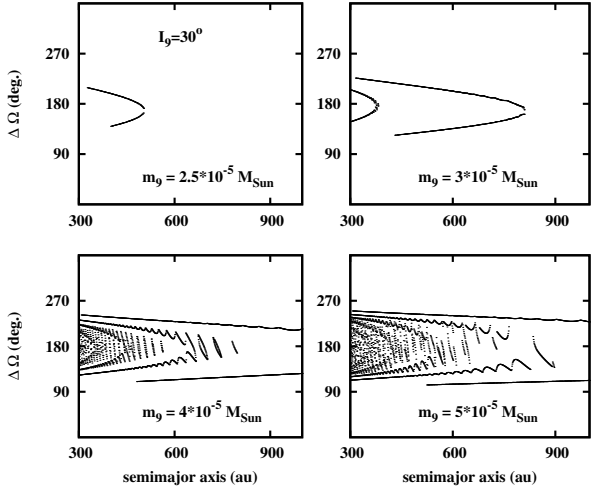}
\caption{Semimajor axis vs. $\Omega$ of pl9s that yield $5.9^{\circ}$ inclination tilt to iv4 with respect to the solar equator, for $I_9=30^{\circ}$.}
\label{ao30}
\end{figure}

\begin{figure}
\includegraphics*[clip=true,scale=0.8]{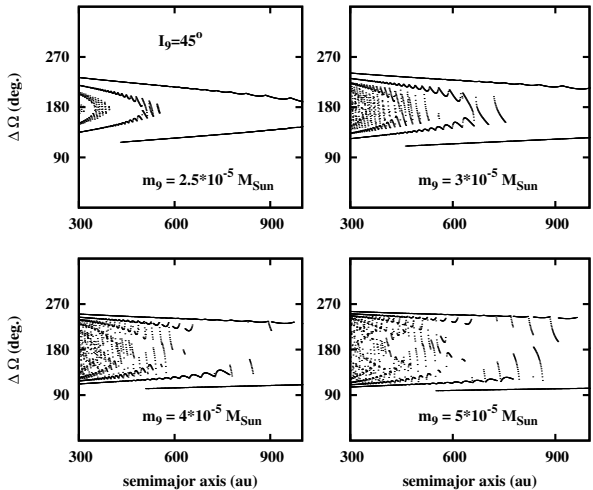}
\caption{Semimajor axis vs. $\Omega$ of pl9s that yield $5.9^{\circ}$ inclination tilt to iv4 with respect to the solar equator, for $I_9=45^{\circ}$}
\label{ao45}
\end{figure}

As iv4 and pl9 orbital planes evolve, their intersections of the solar equator plane defines a difference of longitudes of ascending nodes on that plane. Figures \ref{ao30} and \ref{ao45} show, respectively, for $I_9=30^{\circ}$ and $I_9=45^{\circ}$, pl9 semimajor axis and longitude of the ascending node ($\Omega$) that yield $5.9^{\circ}$ for iv4, on the solar equator plane, with respect to iv4 longitude of the ascending node at $4.5$ Gy. This is also shown for four possible masses of the distant planet. We see that iv4 and pl9  are on average opposed by $180^{\circ}$ on the solar equator. This allows us to compute possible directions for the  longitude of the ascending node of pl9 on the ecliptic plane.

\begin{figure}
\includegraphics*[clip=true,scale=0.8]{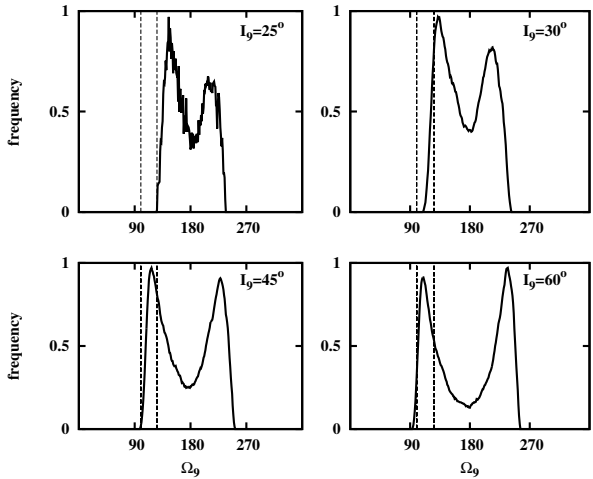}
\caption{Frequency of possible longitudes of the ascending node on the ecliptic that pl9 must have to yield $5.9^{\circ}$ tilt with respect to iv4, for four different inclinations for pl9. Each panel includes masses of pl9 from $2 \times 10^{-5} M_{\odot}$ to $5 \times 10^{-5} M_{\odot}$ with increments of $ 10^{-5} M_{\odot}$. The vertical lines depicts the range of $\Omega$'s determined by \cite{BaBr2016} ($113^{\circ} \pm 13^{\circ}$)}
\label{longs}
\end{figure}

\begin{figure}
\includegraphics*[clip=true,scale=0.8]{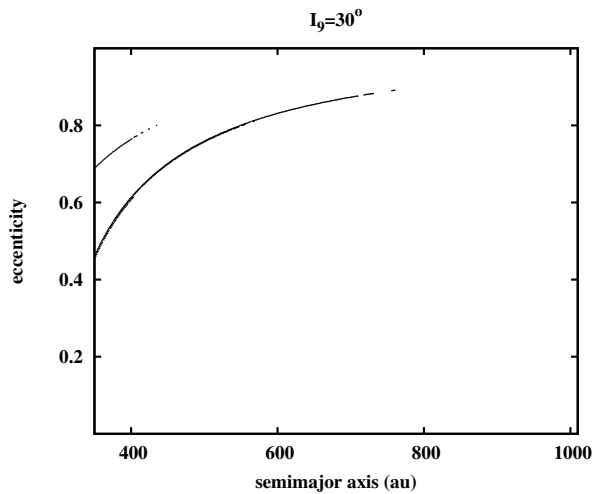}
\caption{Semimajor axis vs. eccentricity of pl9s that yield a tilt of $5.9^{\circ}$ to iv4 with respect to the solar equator, for $I_9=30^{\circ}$, and contrained by the range of longitude of nodes predicted by \cite{BaBr2016}.}
\label{bat}
\end{figure}

Once we know the inclinations and longitudes of the ascending node of pl9 and also the inclination and ascending node of the solar equator with respect to the ecliptic, we can with a couple of rotations compute the longitude of the ascending node of pl9 on the ecliptic.  Figures \ref{ao30} and \ref{ao45} already suggest that pl9's longitude of node will be constrained in a $180^{\circ}$ range. On the ecliptic, Figure \ref{longs} shows the frequency of possible $\Omega$'s of pl9 for four possible $I_9$. All masses of pl9 from $2 \times 10^{-5} M_{\odot}$ to $5 \times 10^{-5} M_{\odot}$ are included in these plots. The vertical lines depicts the range of $\Omega$'s determined by \cite{BaBr2016} ($113^{\circ} \pm 13^{\circ}$). Still for $I_9=30^{\circ}$ Fig. \ref {bat} shows the semimajor axis and eccentricity of pl9 whose $\Omega$ fall inside the range predicted by \cite{BaBr2016}. We see that our approach does not constrain very well the longitude of the ascending node of pl9, but our determination usually includes \cite{BaBr2016} prediction based on the longitude of the ascending nodes of the distant TNO's. Interestingly we have a better match for $I_9>45^{\circ}$. For $I_9=25^{\circ}$ and lower we do not obtain any overlapping with the range of node longitudes from \cite{BaBr2016} work. For $I_9=30^{\circ}$ we have compatibility with \cite{BaBr2016} only for $m_9 \gtrsim 4 \times 10^{-5}M_{\odot} (\sim 13.3 M_{\oplus})$. This seems to indicate that pl9's inclination cannot be much smaller than $30^{\circ}$ and, if so, it needs a mass of the order of $5 \times 10^{-5}M_{\odot} (\sim 17 M_{\oplus})$, somewhat larger than the $10 M_{\oplus}$ usually assumed. This result does not match \cite{malhotra2016}, who give two choices for the pair $I_9$ and $\Omega_9$, which are $(18^{\circ}, 101^{\circ})$ and $(48^{\circ}, 355^{\circ})$. The higher inclination is associated with a longitude of the node that cannot explain the tilt of the giant planets relative to the solar equator (see Fig. \ref{longs}).

It must be noted that when we refer to pl9's inclination we mean its initial inclination with respect to the solar equator plane which coincided with iv4. The final pl9 inclination with respect to the ecliptic which should be compared with current pl9's inclination will vary a little from the initial reference inclination. For instance, for the case where $I_9=30^{\circ}$, the final pl9's inclination with respect to the ecliptic will vary in a range from $27.3^{\circ}$ to $ 37.2^{\circ}$. If we restrict to the $\Omega$'s constrained by \cite{BaBr2016} this range shrinks to $29.7^{\circ}$ to $33.2^{\circ}$. 

\section{Conclusions}

Some ideas had been put forward to possibly explain the inclination of the invariant plane of the known planets relative to the solar equatorial plane \citep{batyginetal2011} but in view of the convincing case presented in \cite{BaBr2016} for the existence of pl9, it is quite natural to suppose that such a tilt was caused by a slow precession of iv4 around the total angular momentum vector of the solar system (including pl9). In this paper we constrained possible masses and orbital elements for pl9 that can account for the present tilt of iv4 with the solar equator. Our results are usually compatible with those of \cite {BaBr2016} and \cite{BrBa2016} though with somewhat larger eccentricities. We also determine a range of possible longitudes of the ascending node for pl9 which often overlaps with the range given in \cite{BaBr2016} except for smaller masses and inclinations of pl9. For instance, for $I_9=30^{\circ}$ we need a mass larger than  $\sim 4 \times 10^{-5} M_{\odot} \sim 13 M_{\oplus}$ to match the range of the longitudes of the ascending node for pl9 proposed by \cite{BaBr2016}.

\section*{Acknowledgments}

R.D. acknowledges support provided by grants \#2015/18682-6 and \#2014/02013-5, S\~ao Paulo Research Foundation (FAPESP) and CAPES. 

\bibliographystyle{apalike}

\bibliography{bib} 

\begin{thebibliography}{}

\bibitem[{Batygin} and {Brown}, 2016]{BaBr2016}
{Batygin}, K. and {Brown}, M.~E. (2016).
\newblock {Evidence for a Distant Giant Planet in the Solar System}.
\newblock {\em \aj}, 151:22.

\bibitem[{Batygin} et~al., 2011]{batyginetal2011}
{Batygin}, K., {Morbidelli}, A., and {Tsiganis}, K. (2011).
\newblock {Formation and evolution of planetary systems in presence of highly
  inclined stellar perturbers}.
\newblock {\em \aap}, 533:A7.

\bibitem[{Beck} and {Giles}, 2005]{beck&giles}
{Beck}, J.~G. and {Giles}, P. (2005).
\newblock {Helioseismic Determination of the Solar Rotation Axis}.
\newblock {\em \apjl}, 621:L153--L156.

\bibitem[{Beust}, 2016]{beust2016}
{Beust}, H. (2016).
\newblock {Orbital clustering of distant Kuiper belt objects by hypothetical
  Planet 9. Secular or resonant?}
\newblock {\em \aap}, 590:L2.

\bibitem[{Brasser} et~al., 2008]{brasseretal2008}
{Brasser}, R., {Duncan}, M.~J., and {Levison}, H.~F. (2008).
\newblock {Embedded star clusters and the formation of the Oort cloud. III.
  Evolution of the inner cloud during the Galactic phase}.
\newblock {\em \icarus}, 196:274--284.

\bibitem[{Brown} and {Batygin}, 2016]{BrBa2016}
{Brown}, M.~E. and {Batygin}, K. (2016).
\newblock {Observational Constraints on the Orbit and Location of Planet Nine
  in the Outer Solar System}.
\newblock {\em \apjl}, 824:L23.

\bibitem[{Fienga} et~al., 2016]{fienga2016}
{Fienga}, A., {Laskar}, J., {Manche}, H., and {Gastineau}, M. (2016).
\newblock {Constraints on the location of a possible 9th planet derived from
  the Cassini data}.
\newblock {\em \aap}, 587:L8.

\bibitem[{Gomes} et~al., 2006]{gomesatal2006}
{Gomes}, R.~S., {Matese}, J.~J., and {Lissauer}, J.~J. (2006).
\newblock {A distant planetary-mass solar companion may have produced distant
  detached objects}.
\newblock {\em \icarus}, 184:589--601.

\bibitem[{Gomes} et~al., 2015]{gomesetal2015}
{Gomes}, R.~S., {Soares}, J.~S., and {Brasser}, R. (2015).
\newblock {The observation of large semi-major axis Centaurs: Testing for the
  signature of a planetary-mass solar companion}.
\newblock {\em \icarus}, 258:37--49.

\bibitem[{Holman} and {Payne}, 2016]{hol-payne2016}
{Holman}, M.~J. and {Payne}, M.~J. (2016).
\newblock {Observational Constraints on Planet Nine: Astrometry of Pluto and
  Other Trans-Neptunian Objects}.
\newblock {\em ArXiv e-prints}.

\bibitem[{Izidoro} et~al., 2015]{izidoroetal2015}
{Izidoro}, A., {Morbidelli}, A., {Raymond}, S.~N., {Hersant}, F., and
  {Pierens}, A. (2015).
\newblock {Accretion of Uranus and Neptune from inward-migrating planetary
  embryos blocked by Jupiter and Saturn}.
\newblock {\em \aap}, 582:A99.

\bibitem[{Malhotra} et~al., 2016]{malhotra2016}
{Malhotra}, R., {Volk}, K., and {Wang}, X. (2016).
\newblock {Corralling a Distant Planet with Extreme Resonant Kuiper Belt
  Objects}.
\newblock {\em \apjl}, 824:L22.

\bibitem[{Murray} and {Dermott}, 1999]{murray-dermott1999}
{Murray}, C.~D. and {Dermott}, S.~F. (1999).
\newblock {\em {Solar system dynamics}}.

\bibitem[{Souami} and {Souchay}, 2012]{souami2012}
{Souami}, D. and {Souchay}, J. (2012).
\newblock {The solar system's invariable plane}.
\newblock {\em \aap}, 543:A133.

\bibitem[{Trujillo} and {Sheppard}, 2014]{trujillo2014}
{Trujillo}, C.~A. and {Sheppard}, S.~S. (2014).
\newblock {A Sedna-like body with a perihelion of 80 astronomical units}.
\newblock {\em \nat}, 507:471--474.

\end{thebibliography}

\end{document}